\documentclass[prl,aps,twocolumn,preprintnumbers,showpacs]{revtex4}
\usepackage{epsfig}
\font\Bbb =msbm10  scaled \magstephalf
\def\id{{\hbox{\Bbb I}}}
\newcommand{\ket}[1]{\big| \, #1 \big\rangle}
\newcommand{\bra}[1]{\left \langle #1 \, \right |}
\newcommand{\proj}[1]{\ket{#1}\bra{#1}}
\newcommand{\braket}[2]{\langle\, #1\,|\,#2\,\rangle}
\newcommand{\outprod}[2]{\ket{#1}\bra{#2}}

\def\opone{\leavevmode\hbox{\small1\kern-3.8pt\normalsize1}}
\newcommand{\tr}[1]{\mbox{Tr} \, #1 }
\newcommand{\vis}{\text{v}}

\begin{document}
\title{Universal Quantum Estimator}
\date{\today}
    \author{Artur K. \surname{Ekert}}
    \author{Carolina \surname{Moura Alves}}\email{carolina.mouraalves@qubit.org}%
    \author{Daniel K. L. \surname{Oi}}%
    \affiliation{Centre for Quantum Computation, Clarendon Laboratory,
      University of Oxford, Parks Road, Oxford OX1 3PU, U.K.}%
    \author{Micha{\l} \surname{Horodecki}}%
    \affiliation{ Institute of Theoretical Physics and Astrophysics,
      University of Gda\'nsk, 80-952 Gda\'nsk, Poland.}
    \author{Pawe{\l} \surname{Horodecki}}%
    \affiliation{Faculty of Applied Physics and Mathematics, Technical
      University of Gda\'nsk, 80-952
      Gda\'nsk, Poland.} %
    \author{L. C. \surname{Kwek}}%
    \affiliation{Department of Natural Sciences, National Institute of
      Education, Nanyang Technological University, 1 Nanyang Walk, Singapore
      637616}

\begin{abstract}
  We present a simple device based on the controlled-SWAP gate that performs
  quantum state tomography. It can also be used to determine maximum and
  minimum eigenvalues, expectation values of arbitrary observables, purity
  estimation as well as characterizing quantum channels. The advantage of this
  scheme is that the architecture is fixed and the task performed is
  determined by the input data.
\end{abstract}


\maketitle

\preprint{Version 8}

One of the the key issues in quantum information is, given an unknown quantum
system, what can we learn about it. In particular, we are concerned not only
with the resources needed (number of identical unknown physical systems), but
also with the complexity of quantum operations required (number of different
devices, networks, etc.), in order to obtain certain information about a
quantum state, characterized by its density matrix $\varrho$. There are many
interesting parameters of $\varrho$ we can determine, such as its maximum and
minimum eigenvalues, its purity or even $\varrho$ itself (state
tomography~\cite{Vogel1989}), but we also can use $\varrho$ to determine
expectation values of arbitrary observables or to characterize unknown quantum
channels. However, this usually involves building separate devices for each
task, or even building different devices for different measurements within the
same task.

In this paper we present a simple, universal device, whose architecture is
fixed but whose behaviour is determined by the choice of input
data~\cite{Nielsenchuang1997} (see also~\cite{Filip2001} for a quantum optical
realization of a similar idea). In fact, with suitable input, we can directly
measure all the properties mentioned before.

\begin{figure}[tbp]
\epsfig{figure=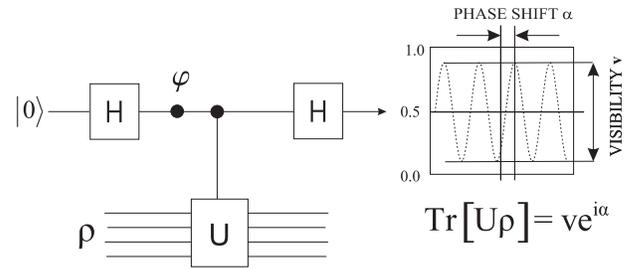,width=0.45\textwidth}
\caption{Both the visibility and the shift of the
  interference patterns of a single qubit (top line) are affected by the
  controlled-$U$ operation.} \label{figint}
\end{figure}

Consider a typical interferometric set-up for a single qubit: Hadamard gate,
phase shift $\varphi$, Hadamard gate, followed by a measurement in the
computational basis. Here and in the following, we borrow terminology from
quantum information science and describe quantum interferometry in terms of
quantum logic gates~\cite{NielsenChuang}. We modify the interferometer by
inserting a controlled-$U$ operation between the Hadamard gates, with its
control on the qubit and with $U$ acting on a quantum system described by some
unknown density operator $\rho$.  We do not assume anything about the form of
$\rho$, it can, for example, describe several entangled or separable
sub-systems. This set-up is shown in Fig.~\ref{figint}. The action of the
controlled-$U$ on $\rho$ modifies the interference pattern by the factor,
\begin{equation}
\tr\rho U = \vis e^{i\alpha},
\label{eqvisi}
\end{equation}
where $\vis$ is the new visibility and $\alpha$ is the shift of the
interference fringes, also known as the Pancharatnam phase~\cite{Pancha56}.

Thus, the observed visibility gives a straightforward way of estimating the
average value of unitary operators $U$ in state $\rho$ and has a variety of
interesting applications. For example, it can be used to measure some
entanglement witnesses $W$, as long as they are unitary operators and the
corresponding controlled-$W$ operations are easy to
implement~\cite{LewensteinKraus2000}. Here, we focus on the applications
related to quantum state state tomography. Clearly the interferometer in
Fig.~\ref{figint} can be used to estimate any $d\times d$ density matrix
$\rho$ by a judicious choice of $d^2-1$ unitary operators $U$ (the basis).
However, this requires building $d^2-1$ networks, each with a different $U$.
We will show that one can build a universal state estimator i.e. a simple
quantum device with a \emph{fixed} architecture, where the state estimation is
performed by modifying \emph{only} the input data.

\begin{figure}[tbp]
  \epsfig{figure=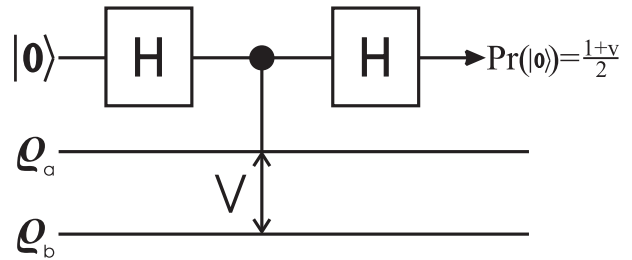,width=0.45\textwidth}
\caption{Universal Quantum Estimator. By judicious choice of input $\varrho_{a}$,
  the state of $\varrho_{b}$ can be determined by examining the change in the
  visibility observed.} \label{figdev}
\end{figure}

In order to do this, let $\rho$ to be composed of two subsystems of dimension
$d$, $\rho =\varrho _{a}\otimes \varrho _{b}$. We will fix our controlled-$U$
to be the controlled-$V$ (SWAP) operator, such that
$V\ket{\alpha}\ket{\beta}=\ket{\beta}\ket{\alpha}$ for any pure states of the
subsystems (Fig.~\ref{figdev}). Let us also introduce the maximally entangled
state $\ket{\phi_+}=1/\sqrt{d}\sum_i\ket{i}\ket{i}$ and let
$P_+=\proj{\phi_+}$. In this case, the modification of the interference
pattern given by~(\ref{eqvisi}) can be written as,
\begin{equation}
\vis=\tr V\left(\varrho _{a}\otimes\varrho _{b}\right)=\tr\varrho_{a}\varrho _{b}.
\end{equation}
Since $\tr \varrho_{a}\varrho _{b}$ is real, we can fix $\varphi=0$ and the
probability of measuring the qubit to be in the state $\ket{0}$ at the output
is related to the visibility by,
\begin{equation}
\vis=2\,\text{Pr}\left(\ket{0}\right)-1.
\end{equation}
If $\rho$ is not separable then, writing $V=P_+^{T_b}$, we obtain $\tr V\rho =
\tr\rho^{T_b}P_+$ (the average value of partially transposed $\rho$ in the
maximally entangled state $P_+$).

Now, let $\varrho_{b}$ be an unknown $d\times d$ density operator. Such an
operator is determined by $d^2-1$ real parameters. In order to estimate matrix
elements of $\varrho_{b}$ in a prescribed orthonormal basis,
$\left\{\ket{n}\right\}$, we proceed as follows: We run the interferometer as
many times as possible (limited by the number of copies of $\varrho_{b}$ at
our disposal) on the input $\proj{\psi}\otimes\varrho_{b}$, where $\ket{\psi}$
is a pure state of our choice. For a fixed $\ket{\psi}$, after several runs we
obtain an estimation of,
\begin{equation}
\vis=\bra{\psi}\varrho_{b}\ket{\psi}.
\end{equation}
The diagonal elements $\bra{n}\varrho_{b}\ket{n}$ can be determined using the
input states $\proj{n}\otimes\varrho_{b}$. The real part of the off-diagonal
element $\bra{n}\varrho_{b}\ket{k}$ can be estimated by choosing
$\ket{\psi}=(\ket{n}+\ket{k})/\sqrt{2}$, and the imaginary part by choosing
$\ket{\psi}=(\ket{n}+i \ket{k})/\sqrt{2}$. In particular, if we want to
estimate the density operator of a qubit, we can choose the pure states,
$\ket{0}$ (spin +$z$), $\left(\ket{0}+\ket{1}\right)/\sqrt{2}$ (spin +$x$) and
$\left(\ket{0}+i \ket{1}\right)/\sqrt{2}$ (spin +$y$).

We can extend the procedure above to estimate expectation values of arbitrary
observables. It can be shown that the mean value of an arbitrary observable
can be reduced to the estimation of a binary two-output POVM \cite{PH}.
Similarly, the mean value $\langle A \rangle_{\varrho_{b}}$ of an arbitrary
observable $A$ in state $\varrho_{b}$ can be measured using the setup in
Fig.~\ref{figdev} with a suitable input $\varrho_{a}$.  We shall apply the
technique utilized in Refs.~\cite{SPA,direct}.  As $A'=\gamma\id+A$ is
positive if $-\gamma$ is the minimum negative eigenvalue of $A$, we can
construct the state $\varrho_{a}=\varrho_{A'}=\frac{A'}{\tr(A')}$ and apply
our interference scheme to the pair $\varrho_{A'}\otimes\varrho_{b}$.  The
visibility gives us the mean value of V (SWAP),
\begin{equation}
\vis=\langle V\rangle_{\varrho_{A'}\otimes\varrho_{b}}=\tr(\varrho_{A'}\varrho_{b}),
\end{equation}
which leads us to the desired value,
\begin{equation}
\langle A\rangle_{\varrho_{b}}\equiv\tr(\varrho_{b} A)=\vis\tr A+\gamma(\vis d-1),
\end{equation}
where $\tr\id=d$.


Other quantities related to $\varrho_{b}$ may be determined by simple
modification of the interferometry scheme. If we have at our disposal two
copies of $\varrho_{b}$ per run, by running the interferometer on the input,
$\rho=\varrho_{b}\otimes\varrho_{b}$, the resulting,
\begin{equation}
\vis=\tr\varrho_{b}^2=\sum_{i}\lambda_{i}^2,
\end{equation}
where $\{\lambda_{i}\}$ are the eigenvalues of $\varrho_{b}$, gives us an
estimate of the purity of $\varrho_{b}$. In the single qubit case, this
measurement allows us to estimate various functionals of $\varrho_{b}$, such
as the length of the Bloch vector,
\begin{equation}
|r|=\sqrt{2\vis-1}.
\end{equation}
Note that the direction, however, is left completely undetermined. The
procedure of estimating eigenvalues and non-linear functionals of
$\varrho_{b}$ can be generalized for larger dimensional systems, but requires
controlled-SHIFT gates~\cite{direct}.


By adapting the input to the interferometer, we can also estimate the extremal
eigenvalues and eigenvectors of $\varrho_{b}$. In this case, the input states
are also of the form $\proj{\psi}\otimes\varrho_{b}$ but we vary $\ket{\psi}$
and search for the minimum and the maximum of
$\vis=\bra{\psi}\varrho_{b}\ket{\psi}$.  This is usually a complicated task as
it involves scanning $2(d-1)$ parameters of $\psi$.  The visibility is related
to the overlap of the reference state, $\ket{\psi}$ and $\varrho_{b}$ by,
\begin{eqnarray}
\vis_{\psi}&=&\tr\left(\proj{\psi}\sum_{i}\lambda_{i}\proj{\eta_{i}}\right)\nonumber\\
 &=&\sum_{i}\lambda_{i}\left|\braket{\psi}{\eta_{i}}\right|^2=\sum_{i}\lambda_{i}p_{i},
\end{eqnarray}
where $\sum_{i}p_{i}=1$. This is a convex sum of the eigenvalues of
$\varrho_{b}$ and is minimized (maximized) when
$\ket{\psi}=\ket{\eta_{min}}\;\left(\ket{\eta_{max}}\right)$. For any
$\ket{\psi}\neq\ket{\eta_{min}}\;\left(\ket{\eta_{max}}\right)$, there exists
a state, $\ket{\psi'}$, in the neighbourhood of $\ket{\psi}$ such that
$\vis_{\psi'}<\vis_{\psi}$ ($\vis_{\psi'}>\vis_{\psi}$) thus this global
optimization problem can be solved using standard iterative methods, such as
steepest decent~\cite{Gill1981}.

Estimation of extremal eigenvalues plays a significant role in the direct
detection of quantum entanglement~\cite{direct} and
distillation~\cite{Horodeckis1997}. In some special cases of two qubits
described by the density operator $\varrho_{b}$ such that at least one of the
qubits is in the maximally mixed state, we can test for the separability of
$\varrho_{b}$ by checking whether the maximal eigenvalue of $\varrho_{b}$ does
not exceed $\frac{1}{2}$ (we shall return to this special case in our
subsequent discussion of the quantum channel tomography).


Let us now turn our attention to characterizing quantum channels. Recall that
a quantum channel is a trace preserving linear map,
$\varrho\rightarrow\Lambda(\varrho)$, which takes quantum states to quantum
states, and whose trivial extensions, $\id_{k}\otimes\Lambda$ do the same,
i.e.  $\Lambda$ is a completely positive map. Using the well known
Jamiolkowski isomorphism~\cite{Jamiolkowski1972} between quantum channels and
bipartite states, quantum channels can be characterized in a simple way. The
single qubit channel capacity and the distillability of entangled states can
be determined by extremal eigenvalue estimation.

\begin{center}
\begin{figure}
  \epsfig{figure=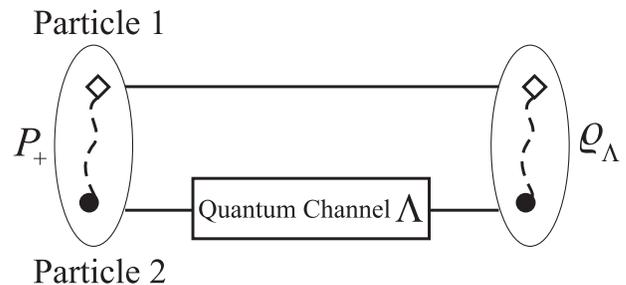,width=0.45\textwidth}
  \caption{A quantum channel is isomorphic with its action on
    half of a maximally entangled bipartite state.}
\label{figchan}
\end{figure}
\end{center}

Suppose we have an unknown quantum channel, $\Lambda$, which we would like to
characterize. The Jamiolkowski isomorphism identifies a quantum channel with
its action on half of a maximally entangled state. The procedure is
thus~(Fig.~\ref{figchan}): We prepare maximally entangled states of two
particles $P_{+}=\frac{1}{d}\sum_{ij}\outprod{i}{j}\otimes\outprod{i}{j}$; We
send one particle through the channel,
\begin{equation}
\label{varrholambda}
P_{+}\rightarrow\left[\id\otimes\Lambda\right] P_{+}=\varrho_{\Lambda};
\end{equation}
We then estimate
\begin{equation}
\varrho_{\Lambda}=\frac{1}{d}\sum_{ij}\outprod{i}{j}\otimes\Lambda\left(\outprod{i}{j}\right),
\end{equation}
which now characterizes $\Lambda$.  We interpret this as $\Lambda$ mapping the
$\outprod{i}{j}^{th}$-element of an input density matrix to the output matrix,
$\Lambda\left(\outprod{i}{j}\right)$. Thus, knowledge of $\varrho_{\Lambda}$
allows us to determine the action of $\Lambda$ on an arbitrary state,
$\varrho\rightarrow\Lambda(\varrho)$.

Now consider the case where $\Lambda$ is a single qubit channel. The channel
capacity, $Q(\Lambda)$ is an important parameter of
$\Lambda$~\cite{Horodeckis2000,huge}.  This is the optimal rate of reliably
sending qubits per use of $\Lambda$. Depending on whether Alice and Bob can
send classical information to each other as an additional resource, one can
consider several capacities, $Q_{C}$ where $C=\o,\leftarrow, \rightarrow,
\leftrightarrow$, corresponding to zero way, one way and two way classical
communication. In general, it is very difficult to calculate the capacity of a
given channel, $\Lambda$. Here we shall provide a simple necessary and
sufficient condition for a one qubit channel to have non-zero two-way
capacity, $Q_{\leftrightarrow}>0$ (obviously a necessary condition for the
other three capacities to be non-zero), which can be determined using our
extremal eigenvalue estimation scheme.

In order to do this, we shall use the state, $\varrho_{\Lambda}$, defined
in~(\ref{varrholambda}).  Clearly, for any channel $\Lambda$,
$\varrho_{\Lambda}$ is maximally mixed when reduced to the subsystem $A$. It
is also true that for any state $\varrho_{\Lambda}$ with a maximally mixed
subsystem $A$ there exists a channel $\Lambda$ which generates
$\varrho_{\Lambda}$ via the formula~(\ref{varrholambda}).  If
$\varrho_{\Lambda}$ is maximally mixed when reduced to both subsystems $A$ and
$B$, then the channel $\Lambda$ is called bistochastic - it maps maximally
mixed states into maximally mixed states.  (Two-qubit states, corresponding to
bistochastic channels, have been completely characterized geometrically as
tetrahedrons in $\mathbf{R}^{3}$~\cite{MRPRA,Oi2001}).


It is known (see~\cite{xor}) that a two qubit state is two-way distillable iff
the operator $\varrho_{A} \otimes I - \varrho_{AB}$ has negative eigenvalue.
Now for states of the type $\varrho_{\Lambda}$ (those are {\it all} states
with $\varrho_{A}=\frac{I}{2}$), this reduces to the requirement that
$\varrho_{\Lambda}$ has maximal eigenvalue greater than $\frac{1}{2}$. This is
also equivalent to $Q_{\leftrightarrow}(\Lambda)>0$, since two-way distillable
entanglement (which is non-zero iff given state is two way distillable) is
simply the lower bound for $Q_{\leftrightarrow}(\Lambda)$~\cite{huge}.

Now we can apply our estimator of the maximal eigenvalue to check whether a
given one-qubit channel has a non-zero two-way capacity. Instead of fully
determining the channel as in the previous section, we simply estimate the
maximum eigenvalue of $\varrho_{\Lambda}$. This scheme can be also be used to
find whether a given two qubit state, with one sub-system maximally entangled,
is two-way distillable.


We have presented a universal quantum estimator, whose action is determined by
the input. The is based on interferometry and the controlled-SWAP operation.
By suitable choice of input states, we are able to perform quantum tomography,
extremal eigenvalue estimation, purity tests and quantum channel
characterization.  Finally let us mention that the controlled-SWAP operation
is a direct generalization of a Fredkin gate~\cite{FT82} and can be
constructed out of simple quantum logic gates~\cite{BBC+95}. This means that
experimental realizations of a universal quantum estimator are within the
reach of quantum technology that is currently being developed.

A.K.E. and L.C.K. would like to acknowledge the financial support provided
under NSTB Grant No. 012-104-0040. P.H. would like to acknowledge the support
of the Polish Committee for Scientific Research and the European Commission.
C.M.A. would like to acknowledge the financial support of Funda{\c c}{\~a}o
para a Ci{\^e}ncia e Tecnologia (Portugal).  D.K.L.O would like to acknowledge
the support of CESG (UK) and QAIP (contract no. IST-1999-11234).

\end{document}